\documentclass{iau} 
\usepackage{graphicx}
\usepackage{amsmath}	% Advanced maths commands
\usepackage{amssymb}	% Extra maths symbols
\usepackage{natbib}

\def\apj{{\it ApJ}}                 % Astrophysical Journal
\def\apjl{{\it ApJ}}                % Astrophysical Journal, Letters
\def\mnras{{\it MNRAS}}             % Monthly Notices of the RAS
\def\aap{{\it A\&A}}                % Astronomy and Astrophysics
\def\icarus{{\it Icarus}}                % Icarus
\def\Science{{\it Science}}

\title[The thermal interplay of gas and dust in protoplanetary disks] %% give here short title %%
{Gas vs dust radial extent in disks: the importance of their thermal interplay}

\author[Stefano Facchini]   %% give here short author list %%
{Stefano Facchini
}

\affiliation{Max-Planck-Institut f\"ur Extraterrestrische Physik, Giessenbachstrasse 1, 85748 Garching, Germany\\ email: {\tt facchini@mpe.mpg.de}\\
}

\pubyear{2017}
\volume{332}  %% insert here IAU Symposium No.
\jname{Astrochemistry VII}
\editors{M. Cunnigham, T. Milar \& Y. Aikawa, eds.}
\begin{document}

\maketitle

\begin{abstract}
A key parameter governing the secular evolution of protoplanetary disks is their outer radius. In this paper, the feedback of realistic dust grain size distributions onto the gas emission is investigated. Models predict that the difference of dust and gas extents as traced by CO is primarily caused by differences in the optical depth of lines vs continuum. The main effect of radial drift is the sharp decrease in the intensity profile at the outer edge. The gas radial extent can easily range within a factor of 2 for models with different turbulence. A combination of grain growth and vertical settling leads to thermal de-coupling between gas and dust at intermediate scale-heights. A proper treatment of the gas thermal structure within dust gaps will be fundamental to disentangle surface density gaps from gas temperature gaps.
\end{abstract}

\firstsection % if your document starts with a section,
\section{Introduction}
Spatially resolved observations of protoplanetary disks in both gas and dust are providing new estimates on their radial extents and surface density profiles \citep[e.g.][]{2014A&A...564A..95P,2017arXiv170608977T,2017arXiv170701499T}. Both quantities are fundamental parameters in any planet formation model \citep{2012A&A...541A..97M}, and together with the disk mass they define the planet formation potential of protoplanetary disks. In particular, the outer radius is a key quantity that could disentangle different evolutionary scenarios of disks, since it is expected to spread outwards in viscous evolution theory \citep{1974MNRAS.168..603L}, whereas spreading is suppressed if disk evolution is driven by magnetic winds \citep{2016ApJ...821...80B}. The radial extent estimates mentioned above are all based on the (sub-)mm continuum brightness distribution, which traces the dust component. Observations of gas tracers, such as rotational transitions of $^{12}$CO, are however showing that the gas seems to extend out to larger radii than the dust. \citet{1998A&A...338L..63D} and \citet{1998A&A...339..467G} suggested that this difference may be due to optical thickness effects, with the gas lines being more optically thick than the (sub-)mm continuum counter-part. However, the steep radial decline in surface brightness at the outer edge of the dust disks has also been interpreted as the smoking gun of radial drift of large dust particles \citep[e.g.][]{2012ApJ...744..162A,2013A&A...557A.133D,2014A&A...564A..95P,2014ApJ...780..153B}.

In this paper, we address this potential difference between gas and dust radial extent taking into account the effects that the dust properties have on the gas temperature and chemistry. Precise estimates of the gas temperature and chemical abundances are needed to correctly predict the molecular emission in the outer regions of protoplanetary disks, which are used in observations to trace outer radii. The grain size distribution of dust particles affects the gas chemistry and temperature, and thus the CO abundance and excitation, in multiple ways. Among the most important effects, the grain size distribution resulting from grain growth, radial drift and vertical settling sets the opacity at far ultraviolet (FUV) wavelengths, and thus the penetration depth of photons capable of dissociating strongly bound molecules as CO. UV radiation and is important for the gas temperature through photoelectric heating. Then, the dust opacities at infrared (IR) wavelengths are important for the gas cooling, which in the upper layers is dominated by IR lines of simple chemical species. The dust surface area also sets the amount of thermal coupling between dust and gas components. Finally, the dust surface area affects the balance between desorption and adsorption of ices onto the surface of dust grains.

A simultaneous modelling of both dust and gas thermal structures in typical protoplanetary disks is performed in the following sections in order to compare the predictions of the dust and gas radial extents when the feedback of realistic grain size distributions of dust particles onto the CO abundance and emission is analysed. The dust properties are computed with state-of-the-art dust evolution codes \citep{2015ApJ...813L..14B}. In order to determine the gas temperatures and chemical abundances of both ice and refractory species, the thermo-chemical code DALI \citep{2012A&A...541A..91B,2013A&A...559A..46B} is used and generalised to consider the underlying dust properties.

\section{Method}
\label{sec:method}

\begin{figure*}
\center
\includegraphics[width=\textwidth]{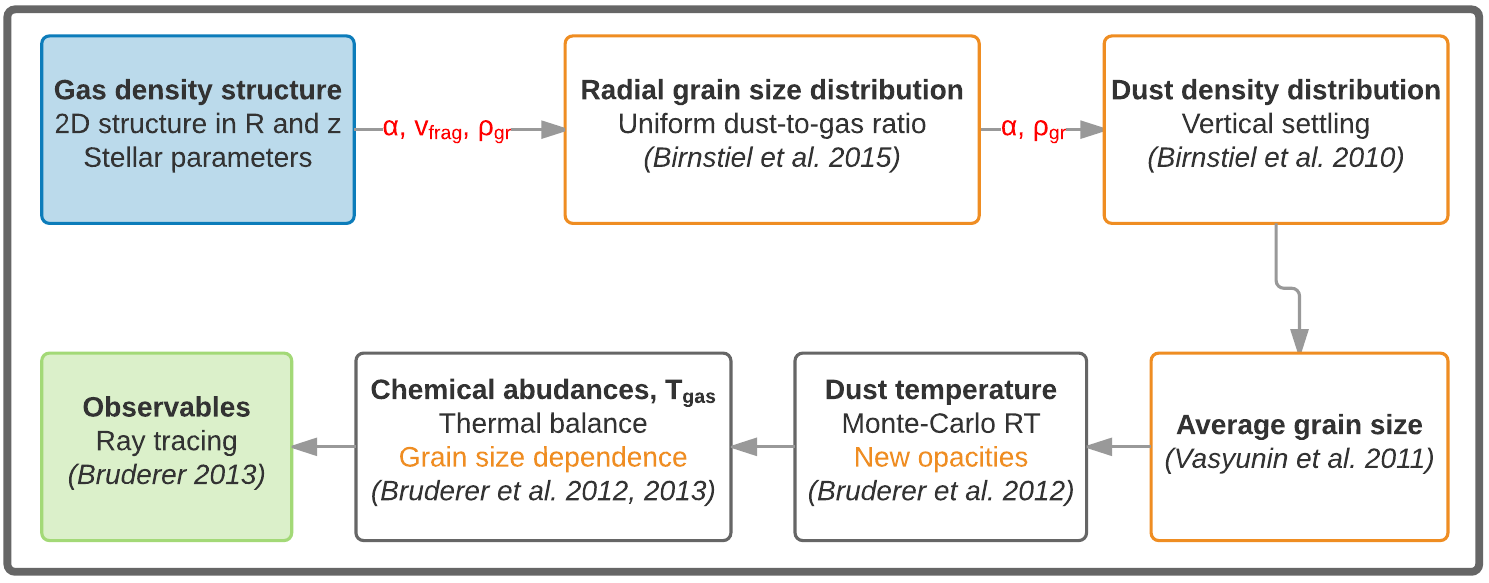}
\caption{Diagram of the new DALI version used in this paper \citep[taken from][]{2017arXiv170506235F}. The blue box shows the input parameters of the models, in particular the disk gas density structure, and stellar parameters. Red parameters indicate physical quantities that are needed to compute the dust properties across the disk. Boxes with orange contours are DALI modules developed for this particular project. The final green box indicates the observables produced with the DALI ray-tracing.
}
\label{fig:diagram}
\end{figure*}

\begin{figure*}
\center
\includegraphics[width=.45\textwidth]{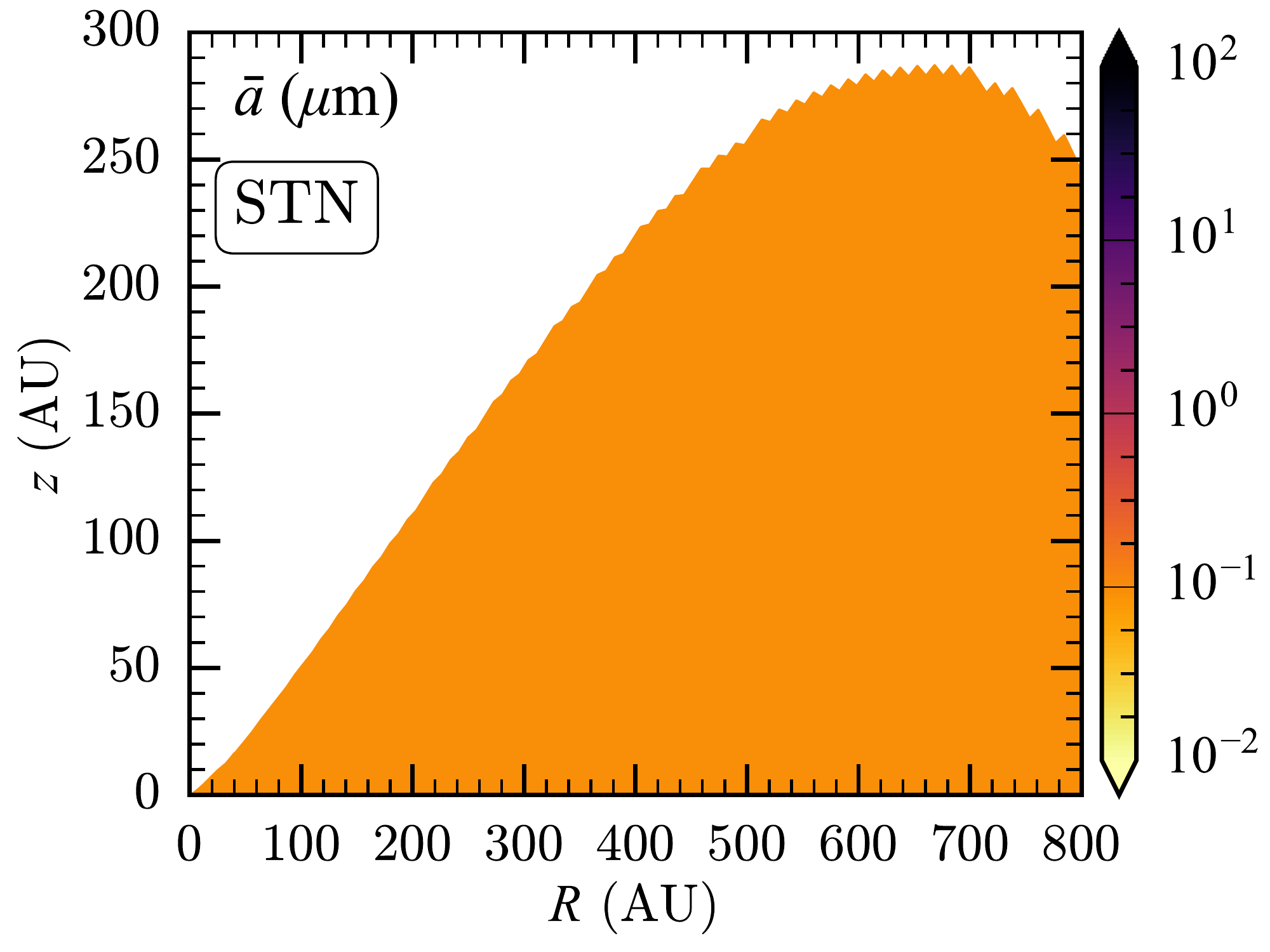}
\includegraphics[width=.45\textwidth]{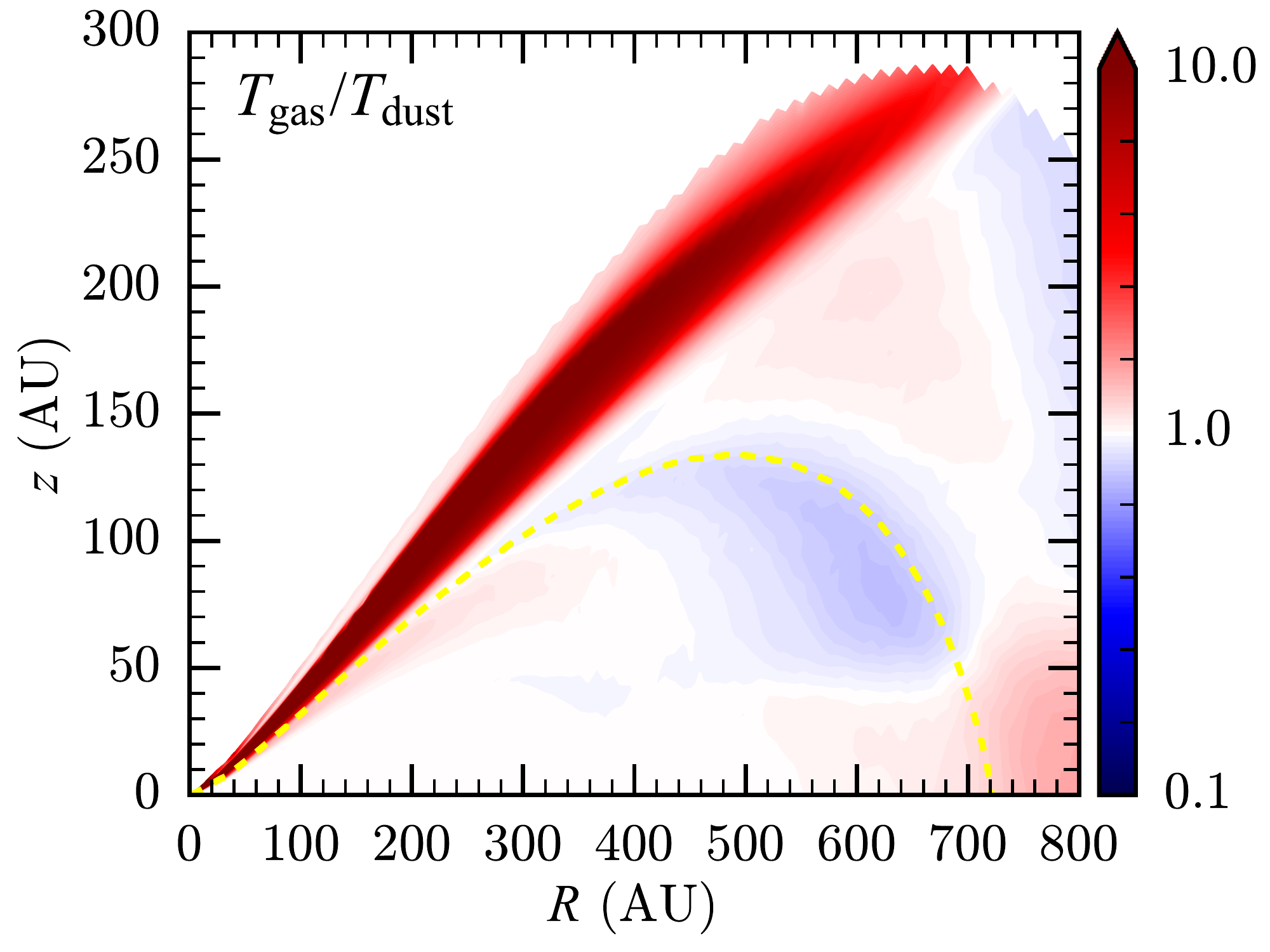}\\
\includegraphics[width=.45\textwidth]{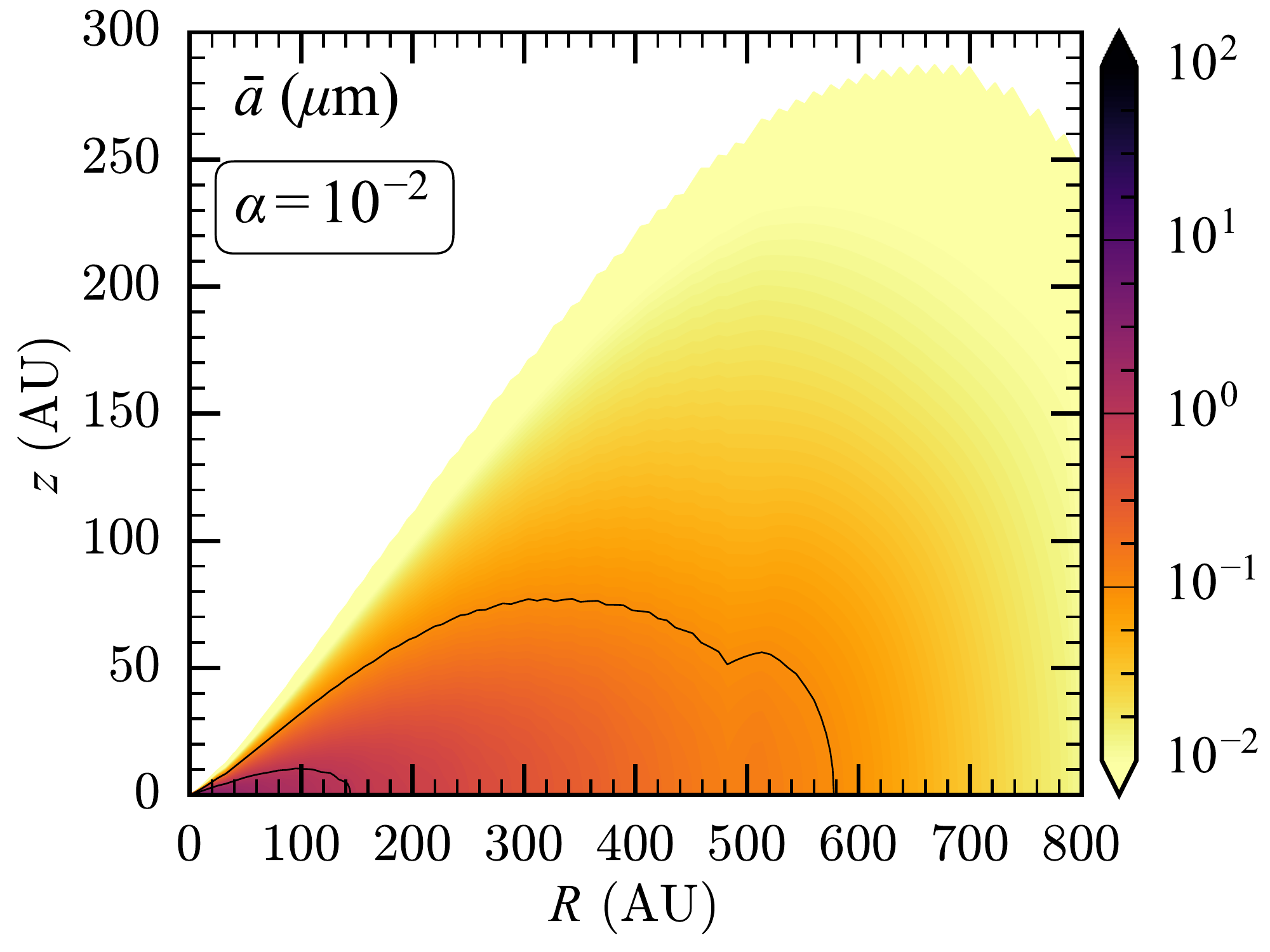}
\includegraphics[width=.45\textwidth]{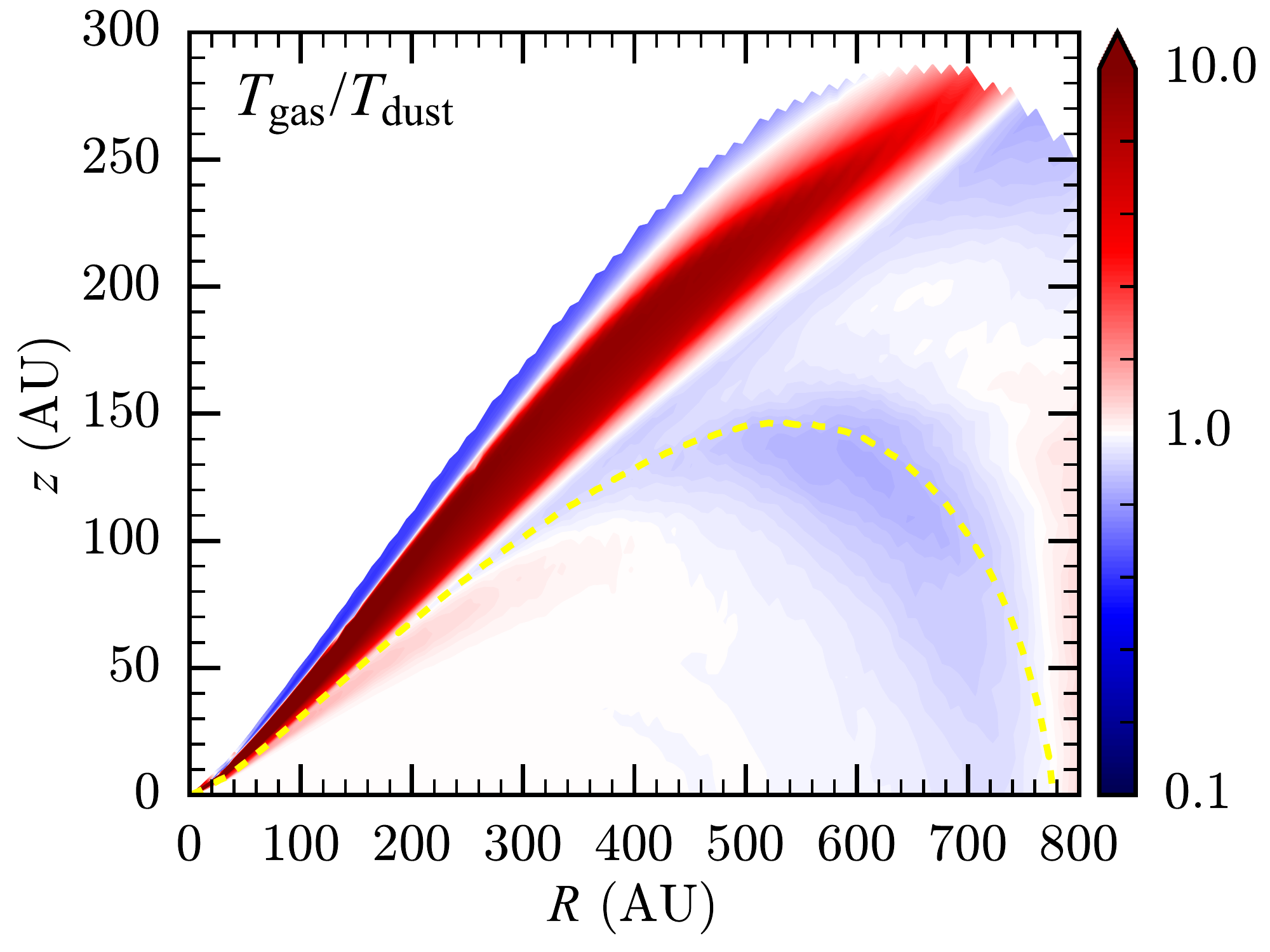}\\
\includegraphics[width=.45\textwidth]{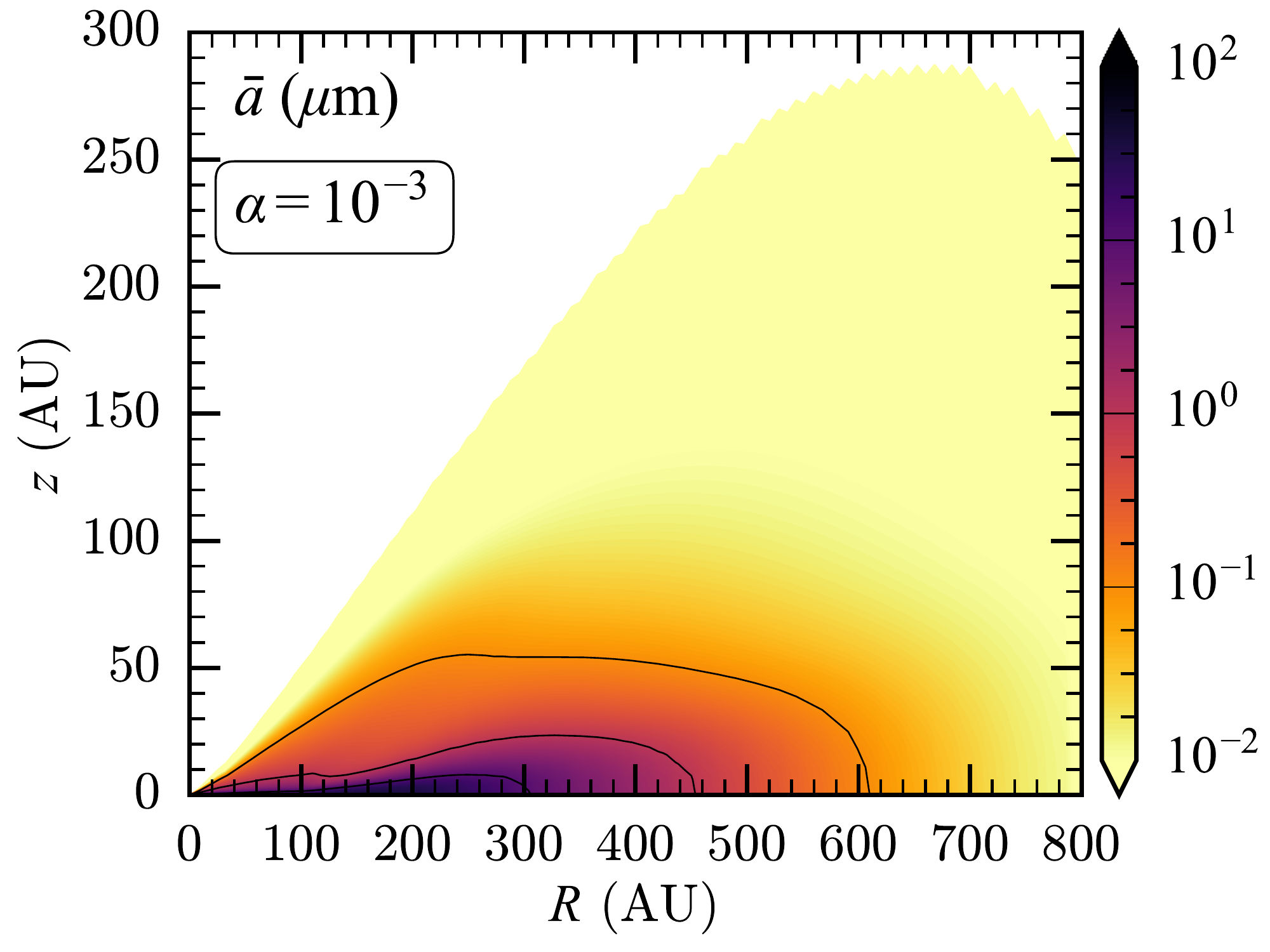}
\includegraphics[width=.45\textwidth]{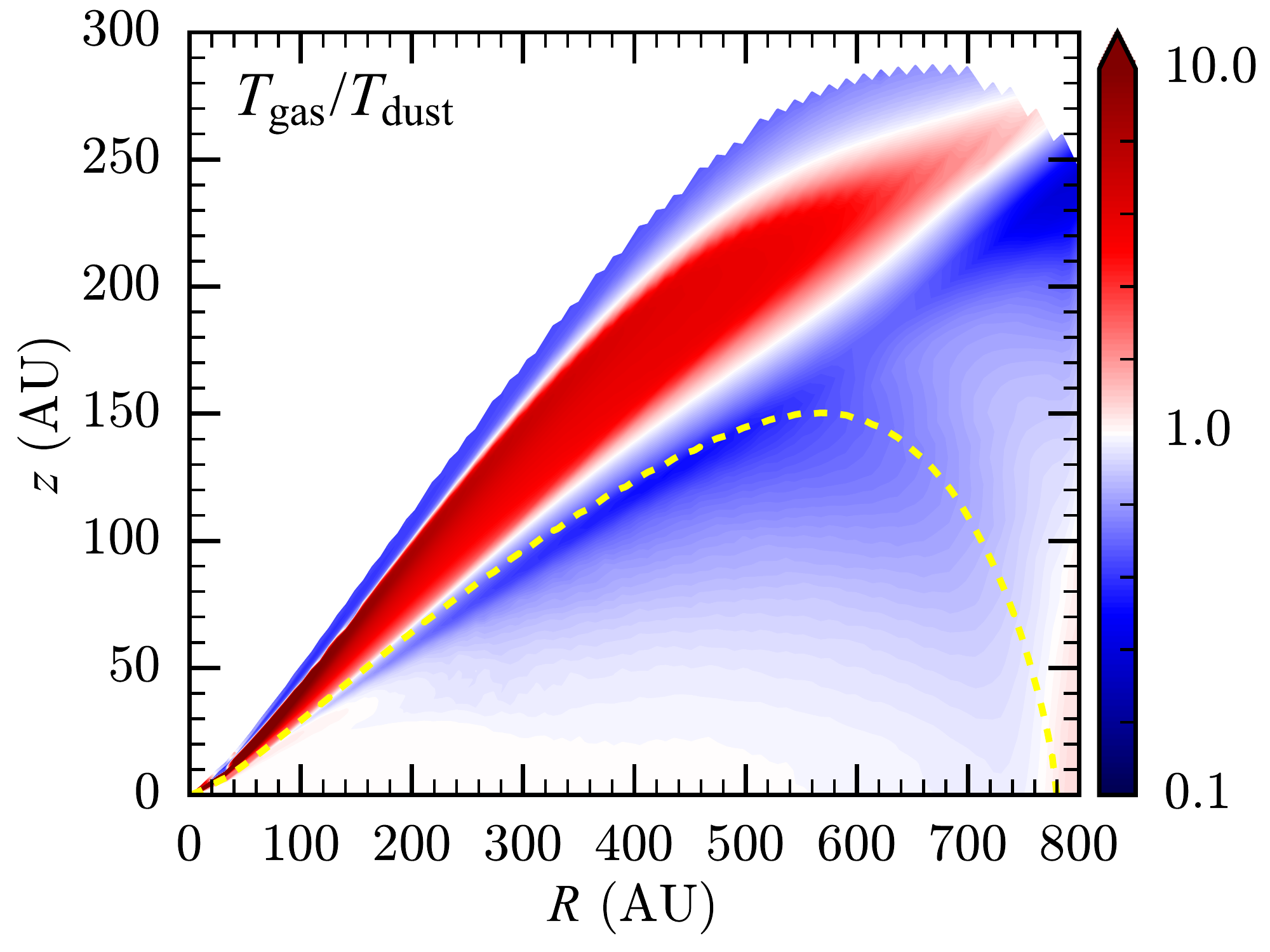}\\
\includegraphics[width=.45\textwidth]{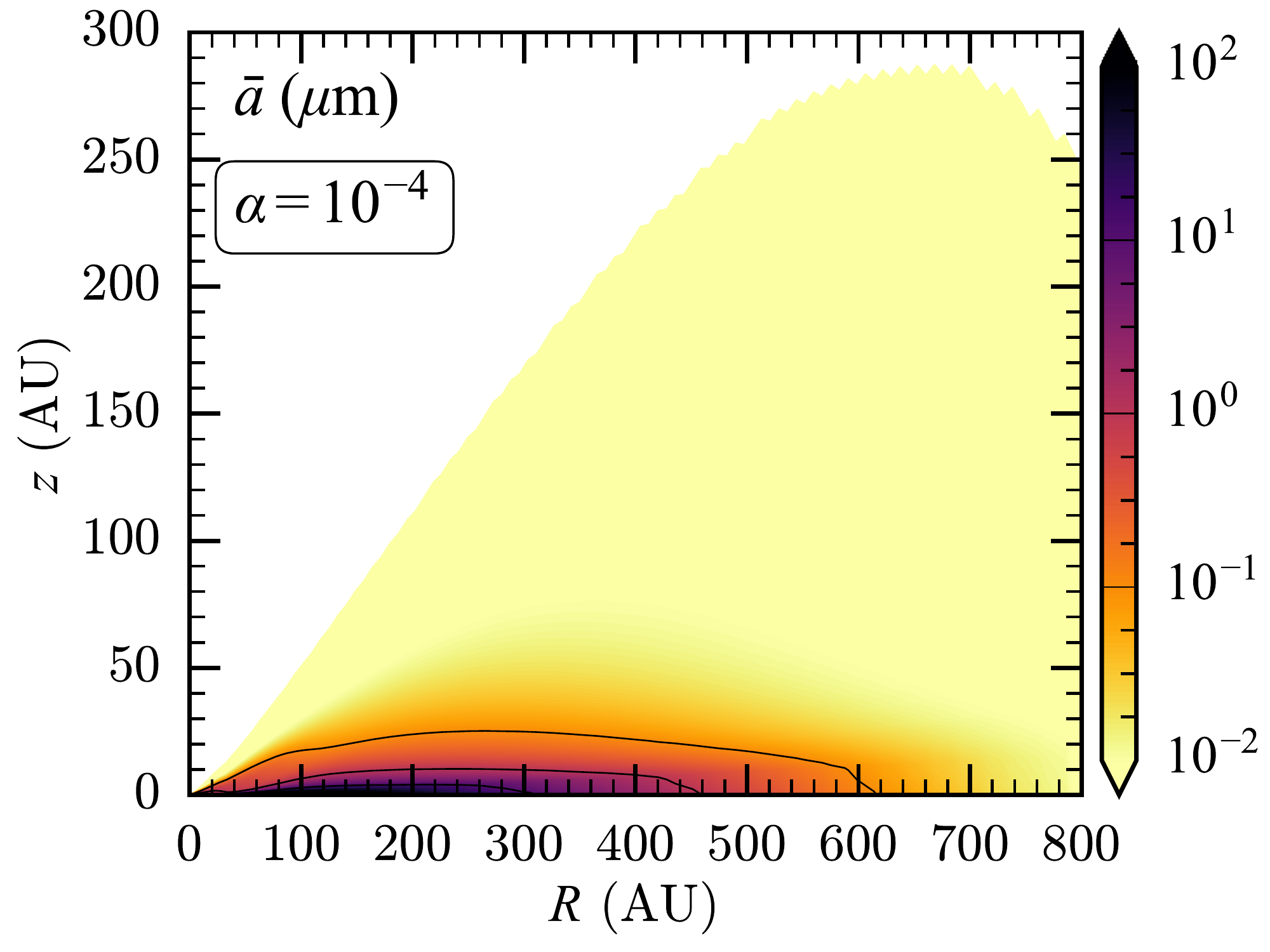}
\includegraphics[width=.45\textwidth]{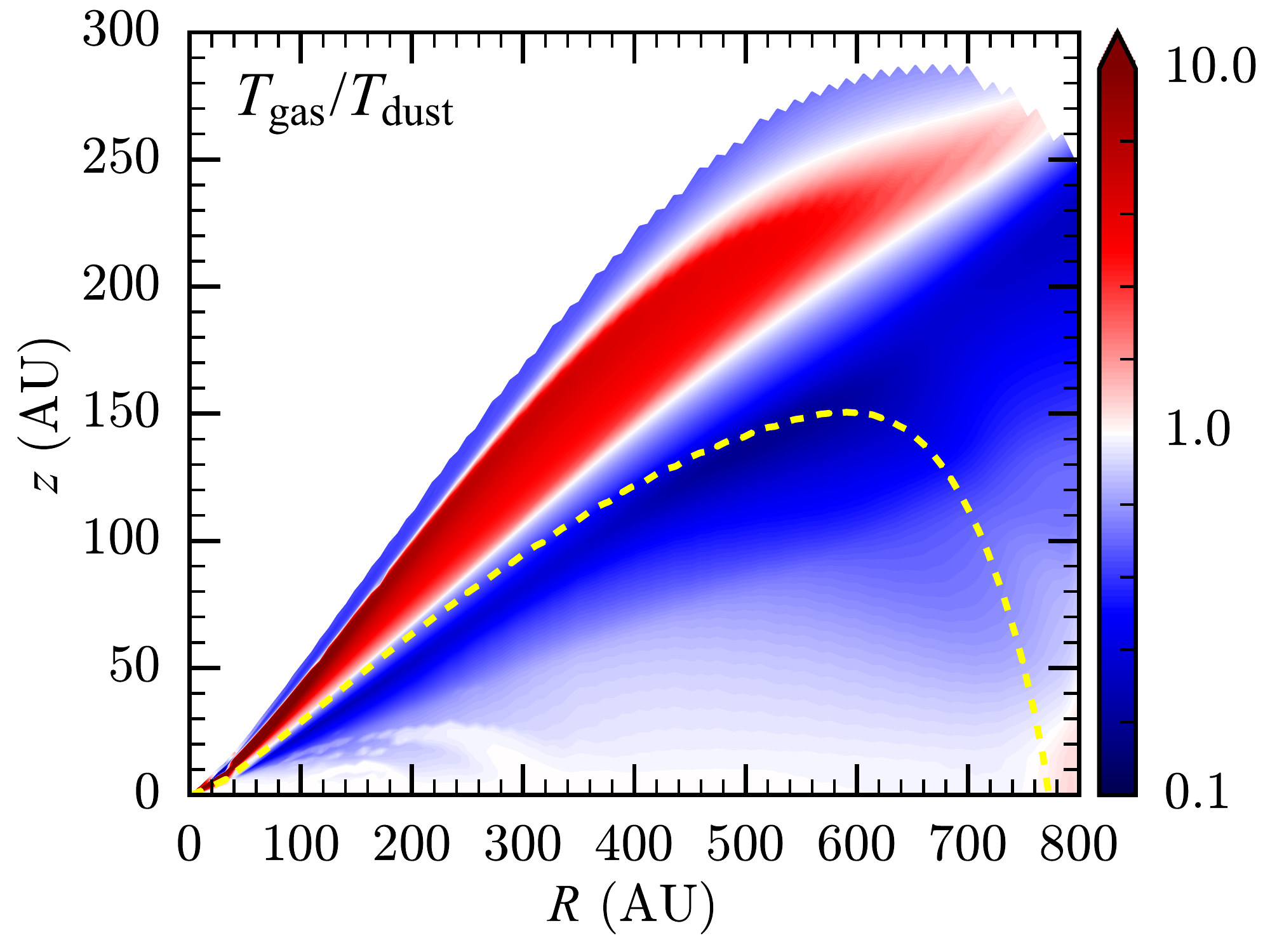}
\caption{Left panels: average dust grain size used in the thermo-chemical calculations for a standard DALI model (STN), and three models with $\alpha=10^{-2}$, $10^{-3}$ and $10^{-4}$, from top to bottom. Note the growth of the average dust size to mm sizes in the disk midplane. Right panels: ratio of gas and dust temperatures for the same models. The yellow dashed lines indicate the $\tau=1$ layer of the $^{12}$CO $J$=3-2 transition. Note that $T_{\rm gas}<T_{\rm dust}$ in the $^{12}$CO emitting layer for low $\alpha$ values. Figure adapted from \citet{2017arXiv170506235F}.
}
\label{fig:summary}
\end{figure*}

The method used in this project is thoroughly described in \citet{2017arXiv170506235F}. Here we present a short summary. The most important addition to previous thermo-chemical models is the calculation of realistic dust properties based on dust evolution models. The main input parameters of these calculations are the gas density structure, which in DALI is two dimensional in $R$ and $z$, the stellar parameters, and the dimensionless turbulent parameter $\alpha$ \citep{1973A&A....24..337S}, which determines both the radial and the vertical dependence of the dust grain size distribution. The gas surface density is parametrized as an exponentially tapered power-law, with a gaussian vertical profile to mimic an isothermal steady-state. The radial grain size distribution is computed using the semi-analytical prescription of \citet{2015ApJ...813L..14B}, where processes such as grain growth, fragmentation, erosion, and radial drift are taken into account. The calculations are performed using 250 bins in grain size, logarithmically sampled between $50\,$\AA\ and $10$\,m. The dust-to-gas ratio is kept fixed (at a $0.01$ value) in the whole radial grid. Thus, radial drift is taken into account in the determination of the largest particle size attainable in the disk, when fragmentation does not dominate, but it is not self-consistently considered to compute the evolution of the dust-to-gas ratio. The radial grain size distribution is then used to compute the vertical segregation of dust particles, based on the steady-state advection diffusion equation of vertical settling \citep[e.g.][]{1995Icar..114..237D}. With these calculations, mass-averaged opacities can be calculated in every point in the disk, and the dust temperature is determined with the DALI continuum radiative transfer.

As for the thermo-chemistry, DALI self-consistently computes the gas temperature and chemical abundances, based on a reduced chemical network. Both in the thermal balance and in the chemistry there are processes that depend on the dust surface area available, in particular:

\begin{itemize}
\item freeze-out and desorption
\item photo-processes on dust grains
\item gas-grain collisions;
\item H$_2$ formation
\item hydrogenation
\end{itemize}
Obviously, there is also a clear dependence on the dust temperature and the continuum intensity spectrum. The code has been upgraded to take into account the dependence of all these effects on the dust surface area. From the density distribution of the $250$ grain sizes, the average dust surface area is computed in every point of the disk using the prescription by \citet{2011ApJ...727...76V}. 

\begin{figure*}
\center
\includegraphics[width=.49\textwidth]{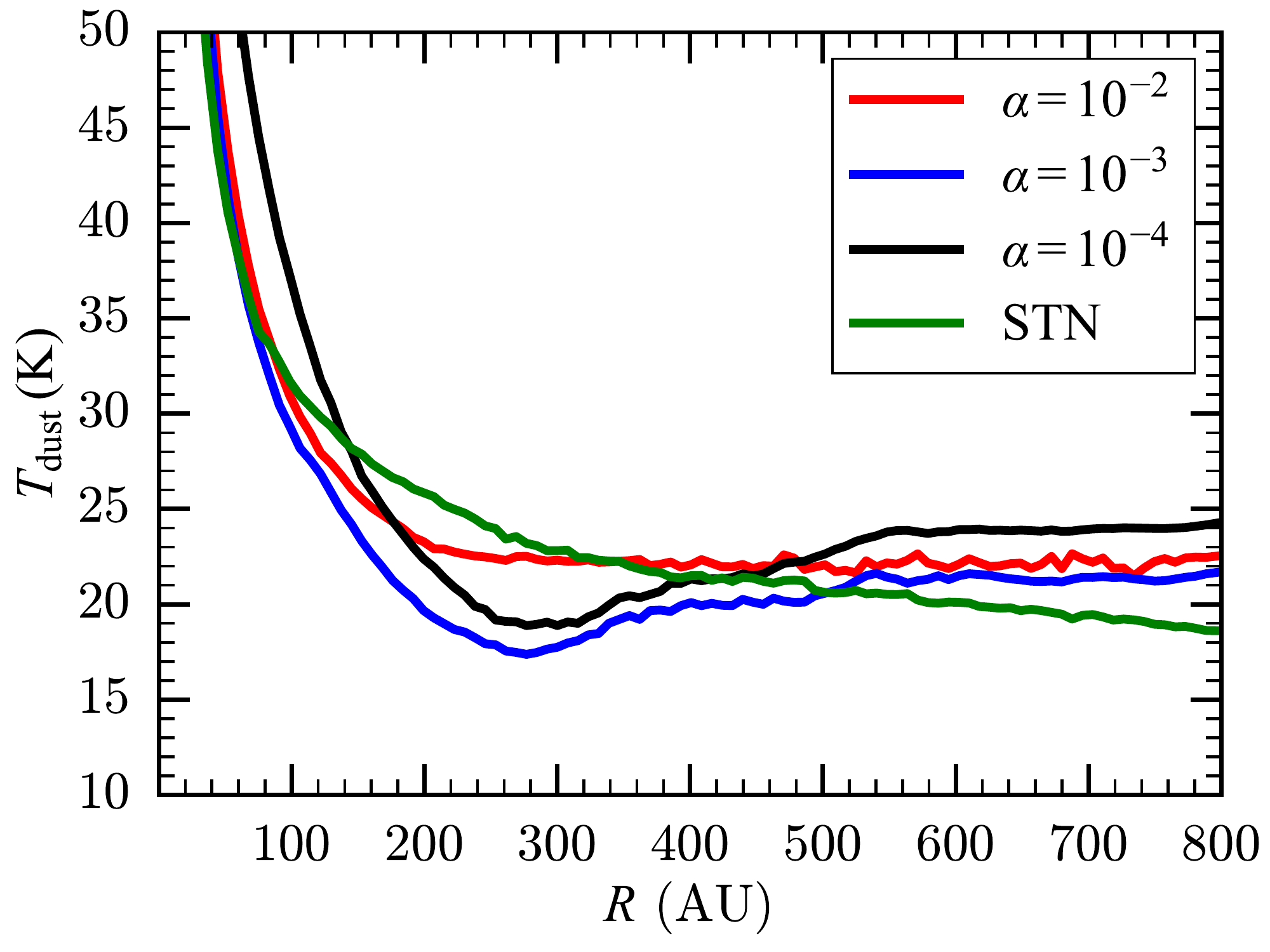}
\includegraphics[width=.49\textwidth]{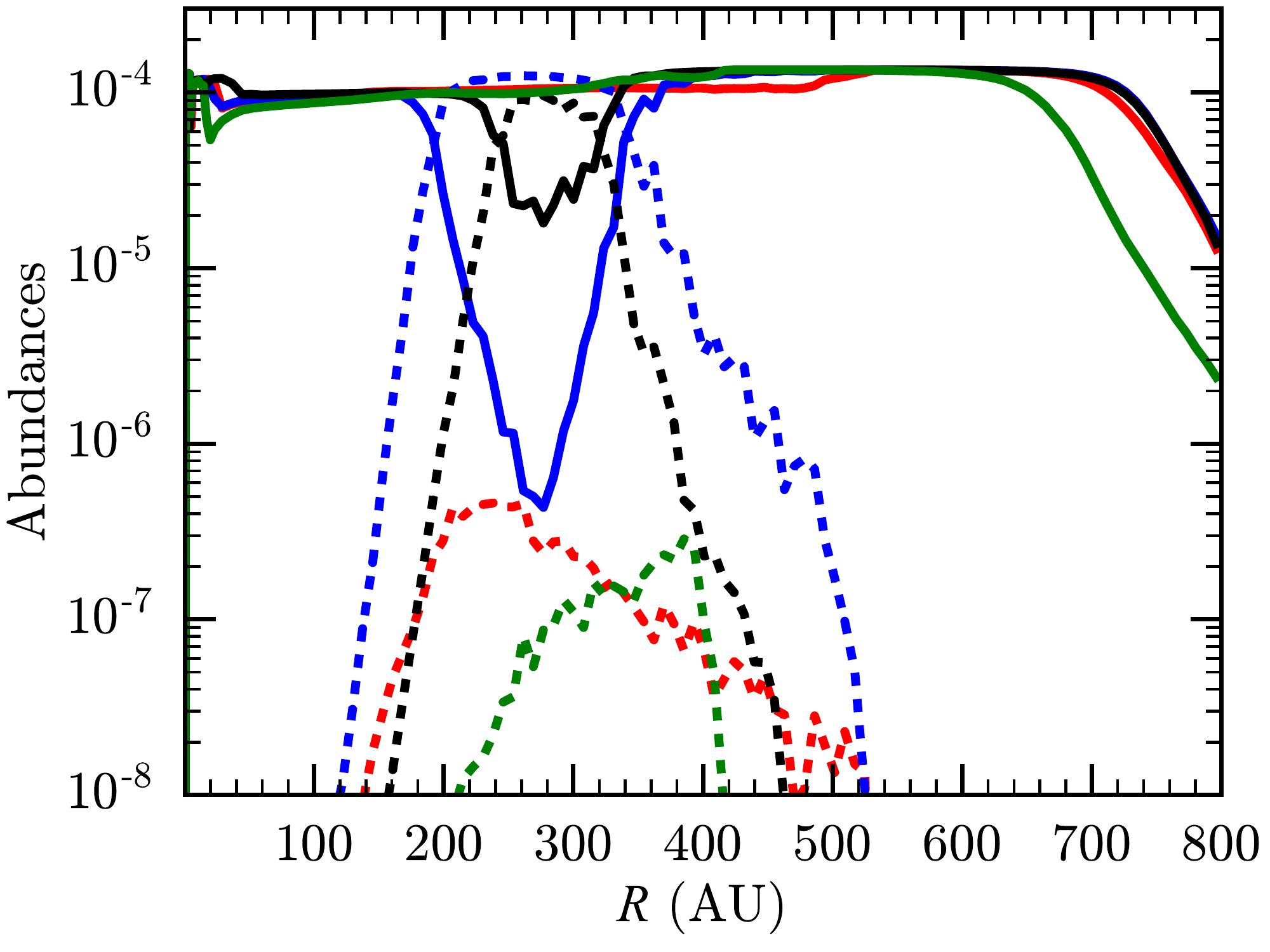}
\caption{Left panel: dust midplane temperatures of different models. Right panel: volatile and solid CO abundances in the disk midplane for the same models. Solid lines represent gas phase CO, dashed lines indicate CO ice. For low turbulent values ($\alpha\lesssim 10^{-3}$) the disk midplane can witness a thermal inversion in its outer regions, due to a combination of grain growth and vertical settling. Figure adapted from \citet{2017arXiv170506235F}.
}
\label{fig:snowline}
\end{figure*}

The results of the models containing grain growth and vertical settling are compared against the output of a model where the dust properties are more simply parametrised. In the next sections this model will be referred to as ``STN''. All the models shown in this paper have the same gas density structure and stellar parameters, which are taken as the best representative values of the disk HD163296 from \citet{2013A&A...557A.133D}. The differences that are observed between the models are due to the different dust grain size distributions across the disk.

\section{Dust and gas temperatures}

Grain growth and vertical settling have a significant effect both on the dust and on the gas temperatures. The left panels of Figure \ref{fig:summary} show the average grain size that is obtained across disks where different turbulent parameters have been used, ranging between $\alpha=10^{-2}$ and $10^{-4}$. Comparison with the typical value of $0.1\,\mu$m used in the chemical calculations of disks is also portrayed. As turbulence decreases, the dust becomes more severely settled towards the disk midplane. Moreover, the maximum grain size reached in the disk midplane at a given radius is larger for lower turbulence, since turbulent velocities of dust particles become lower, and thus grains are less prone to fragment.

Severe vertical settling can lead to the peculiar effect of having a region where gas is colder than the dust particles at intermediate scale heights (see Figure \ref{fig:summary}, right panels). In the upper layers of the disk, the gas is usually much hotter than the dust, in particular in the region that is still optically thin to FUV wavelengths, since most of the gas heating in these regions comes from photoelectric effect on polyciclic aromatic hydrocarbons (PAHs). When significant vertical settling of dust particles occurs, there are two effects that come into play: there is a region in the intermediate layers that is optically thick to UV wavelengths, but optically thin to IR wavelengths, so most of the cooling lines can still escape. Moreover, in this same region the dust densities are not high enough to lead to thermal coupling between the gas and the dust components via gas-grain collisions. The combination of the two effects can make the gas colder than the dust component. Interestingly, on the same panels in Figure \ref{fig:summary}, the $\tau=1$ layer of the $^{12}$CO $J$=3-2 transition is overlaid, and it crosses exactly the region where the thermal de-coupling occurs.

In the disk midplane, when turbulence is low, there can be a radial range where the disk is self-shadowed, in particular when the radial gradient of the maximum grain size is steeper than the radial gradient of the gas surface density. However, in the disk outer regions, the dust particles are small enough that they are stirred up to high altitudes very easily. These two effects combined lead to a thermal inversion in the disk midplane (see Figure \ref{fig:snowline}, left panel), with a dust temperature that can radially increase in the outer regions of the disk. For a warm disk, as the one analysed in this paper, this inversion can happen close to the sublimation temperature of CO, at  $\sim20-25\,$K. The right panel of Figure \ref{fig:snowline} shows that for the $\alpha=10^{-3}$ and $10^{-4}$ models, CO can freeze-out at the lowest temperatures of the respective models, but then is desorbed back into the gas phase in the outer regions of the disk, where the dust temperatures are higher than the CO sublimation temperature. In these calculations, a CO binding energy of $855\,$K was assumed, typical of CO pure ice \citep[e.g.][]{2003ApJ...583.1058C}. In the models presented here, this second desorption front is due to a temperature inversion, not to an increase penetration depth of UV photons, as in the models by \citet{2016ApJ...816L..21C}. Interestingly, in both scenarios radial drift plays an important role in predicting this trend, which has  been observed (or inferred) in recent ALMA images \citep[e.g.][]{2016ApJ...823...91S,2016ApJ...823L..18H,2017arXiv170602608D,2017arXiv170706475S}. 

\begin{figure*}
\center
\includegraphics[width=.49\textwidth]{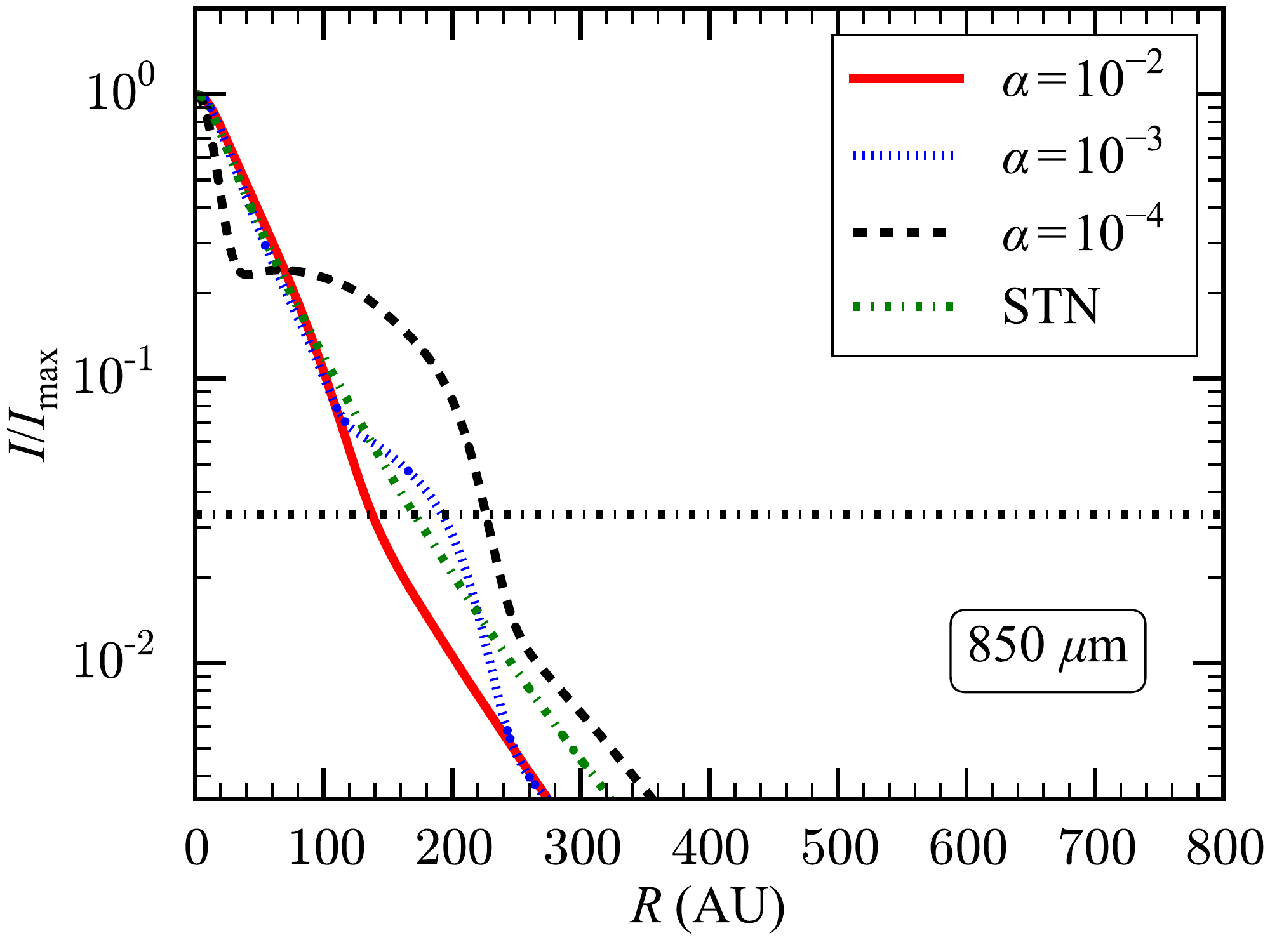}\\
\includegraphics[width=.49\textwidth]{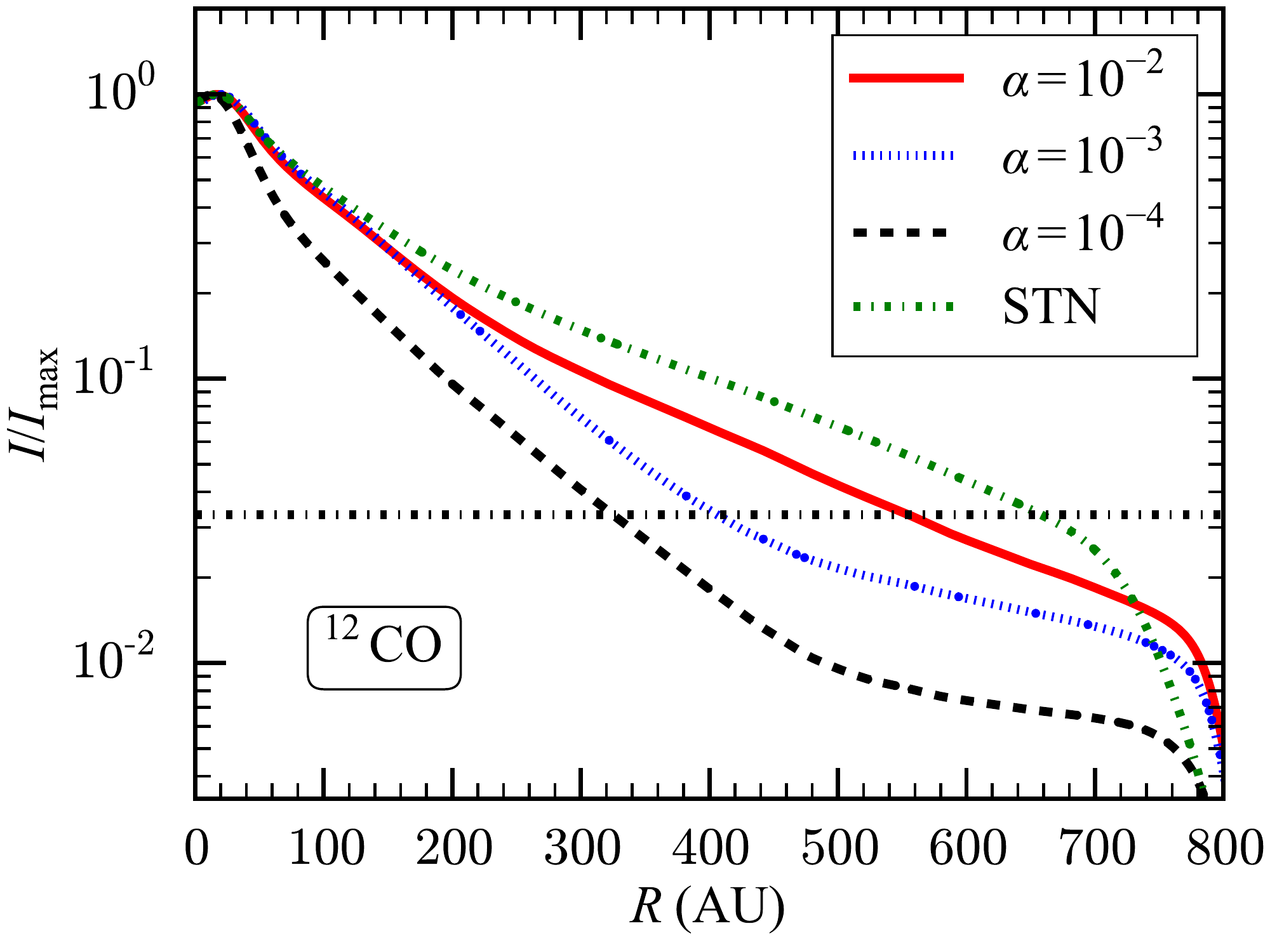}
\includegraphics[width=.49\textwidth]{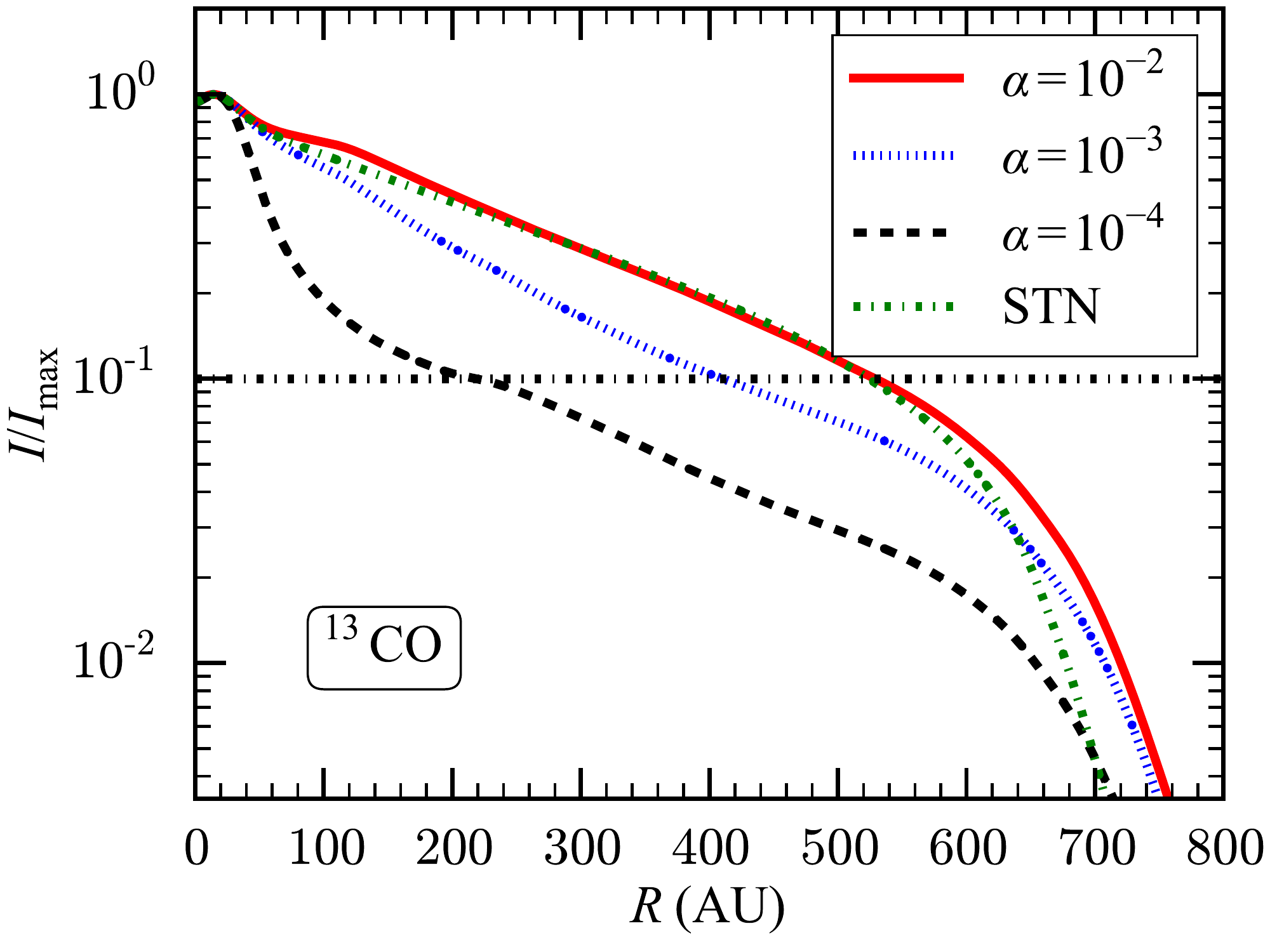}
\caption{Normalised radial intensity profiles of the $850\,\mu$m continuum (top panel), $^{12}$CO $J$=3-2 line, and $^{13}$CO $J$=3-2 line. The dashed-dotted black line indicates a dynamic range of $30$ for both the continuum and $^{12}$CO, and a dynamic range of $10$ for $^{13}$CO. Note that the underlying gas surface density is the same in all models. Panels adapted from \citet{2017arXiv170506235F}.
}
\label{fig:profiles}
\end{figure*}

\begin{figure*}
\center
\includegraphics[width=.49\textwidth]{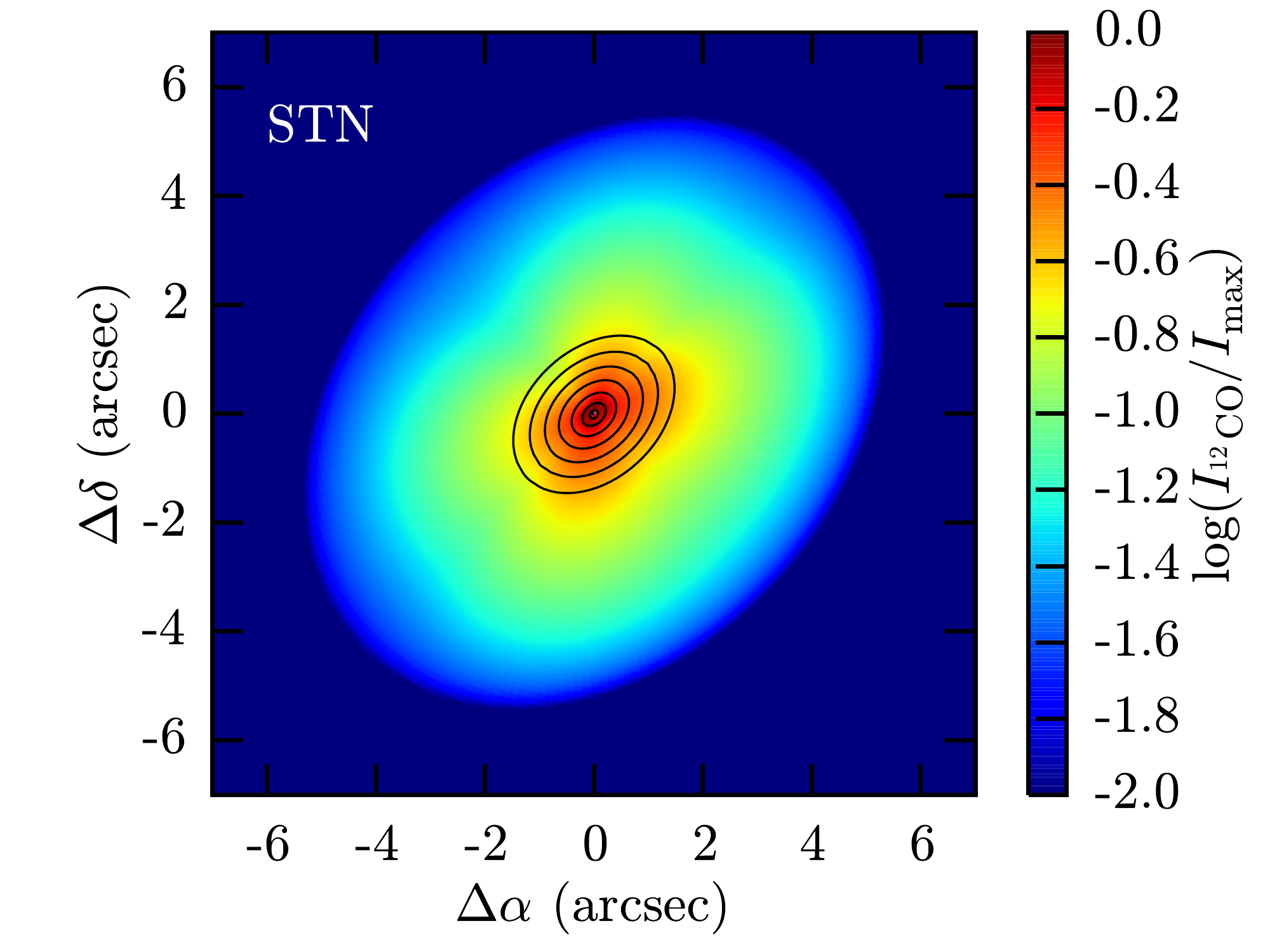}
\includegraphics[width=.49\textwidth]{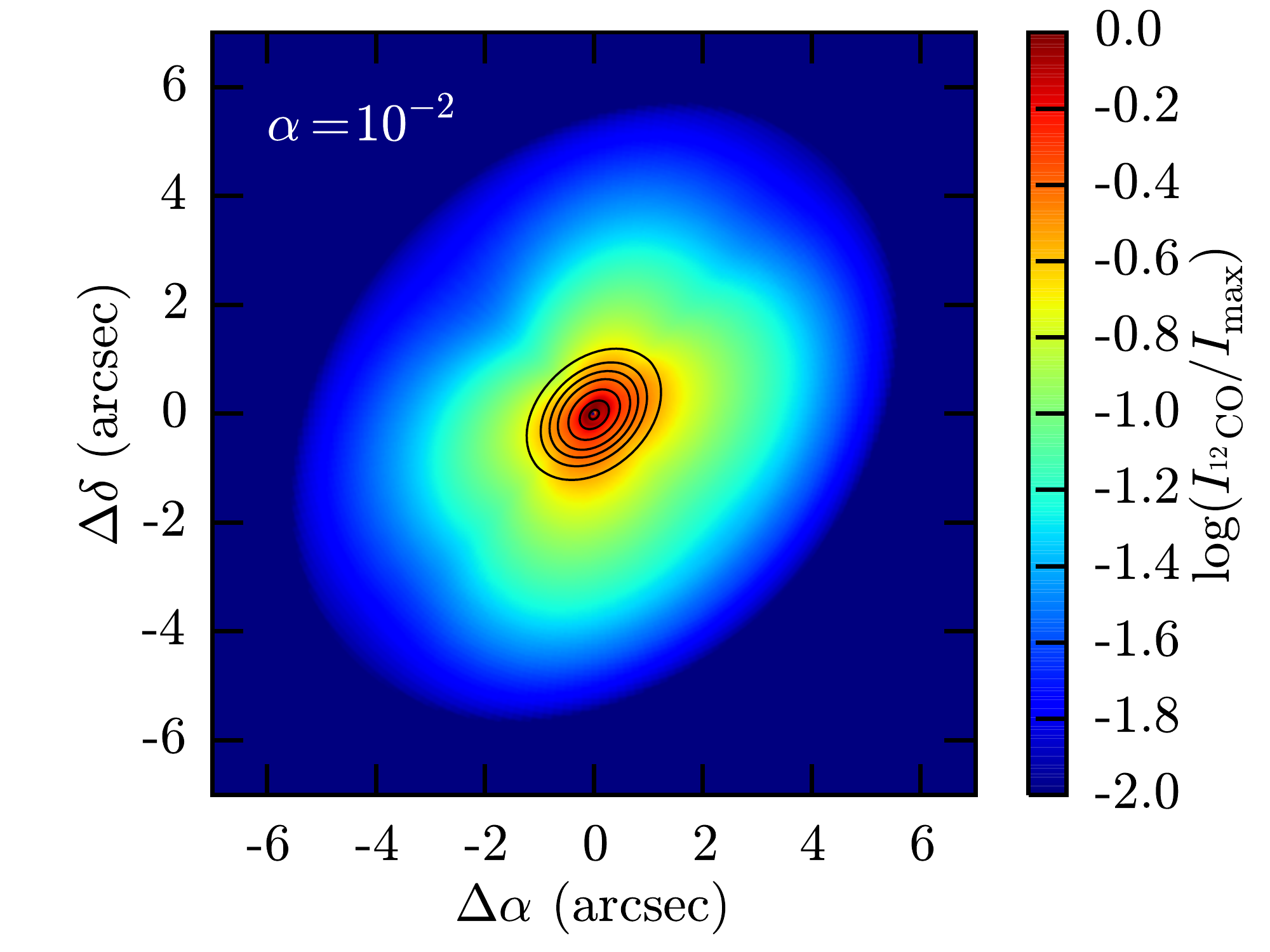}\\
\includegraphics[width=.49\textwidth]{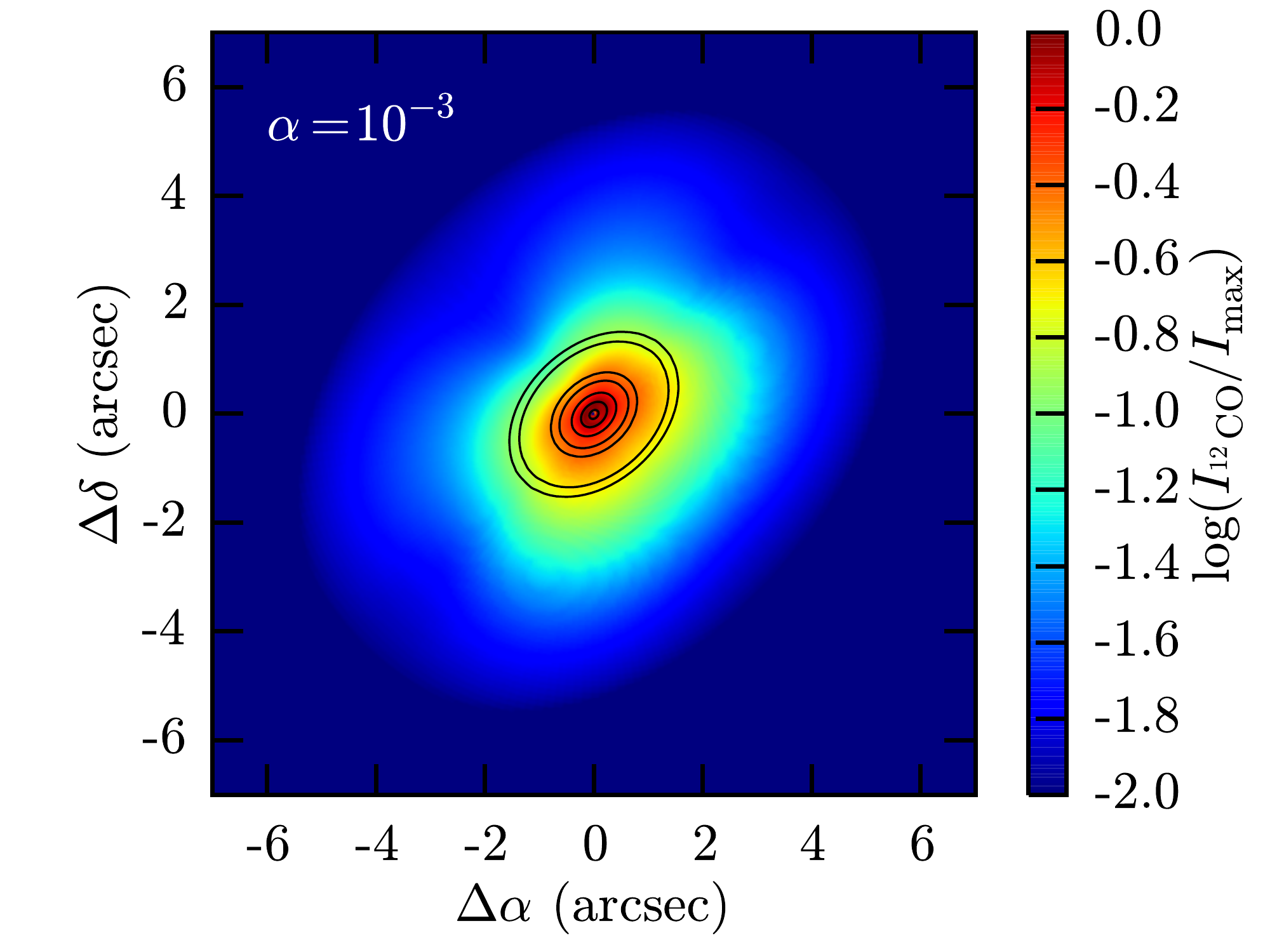}
\includegraphics[width=.49\textwidth]{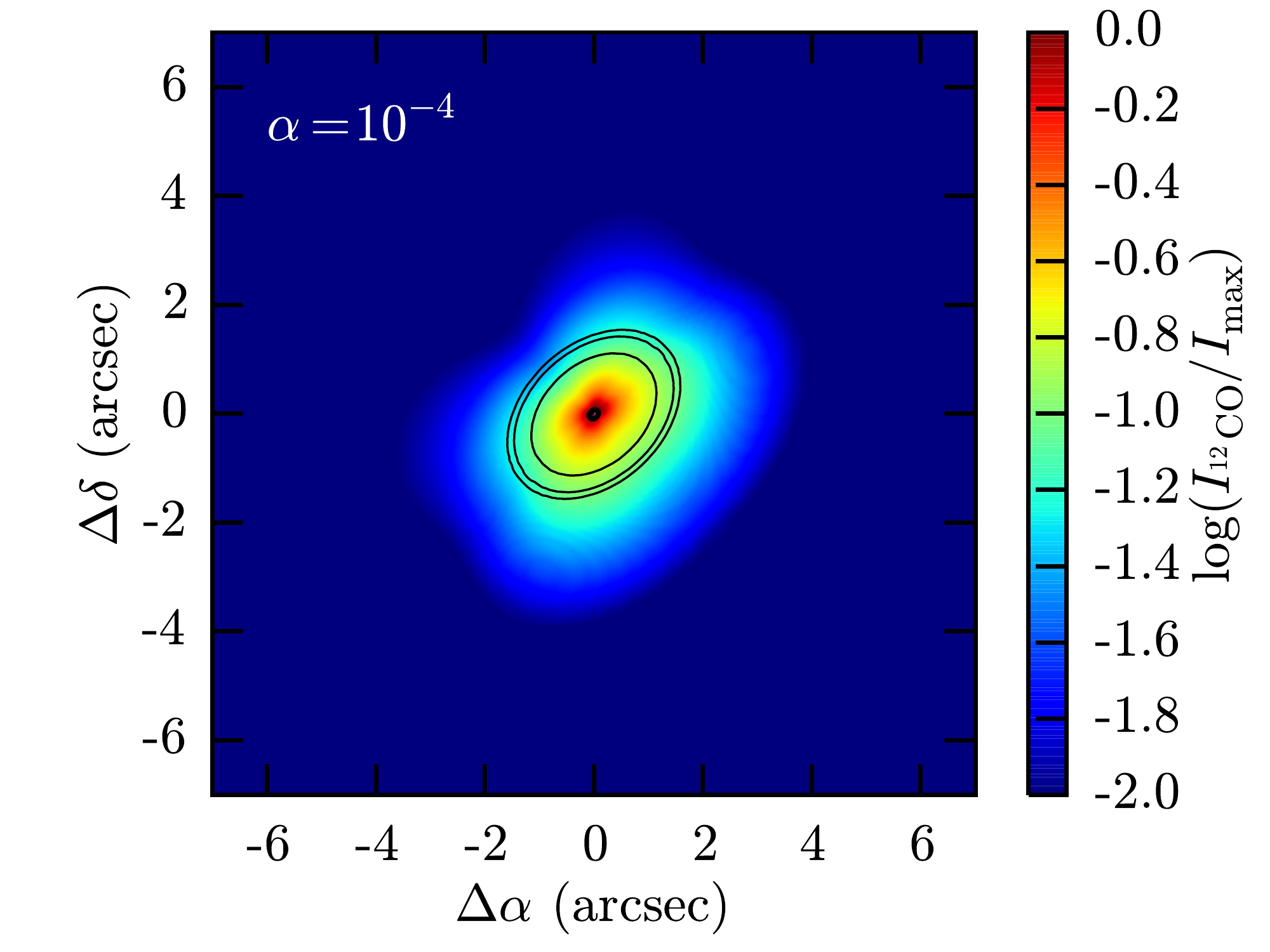}\\
\caption{Normalized intensity maps of the $^{12}$CO $J$=3-2 line for all models. Contours represent the peak normalized continuum levels at $850\,\mu$m between 1/2 and 1/128, logarithmically sampled. Panels from \citet{2017arXiv170506235F}.
}
\label{fig:images}
\end{figure*}

\section{Impact on outer radius estimates}
\label{sec:radius}

It has been shown that the gas temperatures can be significantly affected by grain growth, vertical settling and radial drift. The gas temperatures are important in determining the radial intensity profiles of significantly optically thick and bright molecules, which are the more natural lines to use when trying to determine the radial extent of the gas component of a disk out to the very cold regions. If the radial extent of a disk is defined as the radius where the radial intensity profile reaches a specific value (that can be set by the sensitivity of the observation, or as the radius at which the intensity is lower than the peak by some fixed amount), the thermal coupling between gas and dust can play a significant role. Figure \ref{fig:profiles} shows the radial intensity profiles of the models presented in Section \ref{sec:method}, for sub-mm thermal continuum at $850\,\mu$m, for the $^{12}$CO $J$=3-2 line, and for the $^{13}$CO $J$=3-2 line. At a fixed peak intensity ratio, the estimated radial extent obtained from both $^{12}$CO and $^{13}$CO has a significant dependence on turbulence. Since the gas density structure is exactly the same for all the models, the difference is due to the radial gas temperature profile along the $\tau=1$ layer of the lines being steeper for lower turbulence values. Thus, as the layers where the lines originate become colder, the line intensities become lower in the outer regions of the disk, and the disk looks more compact with a steeper surface density profile (see Figure \ref{fig:images}).

The radial extent as probed by the submm continuum is not significantly affected by grain growth and radial drift. The top panel of Figure \ref{fig:profiles} shows the radial intensity profile at $850\,\mu$m, and the difference between the models including grain growth and radial drift and a model with a uniform grain size distribution (the STN model) is not very significant. However, for low turbulent values, the steep radial gradient in the maximum grain size attained in the disk midplane makes the outer regions of the intensity profile very steep. In these models, this is not due to a radial variation in the dust-to-gas ratio \citep[which can have a significant effect, not considered here, as shown in][]{2014ApJ...780..153B}, which is kept fixed, but to the radial gradient of the grain size distribution, a direct consequence of radial drift. Thus, the difference of the radial extent of the continuum and line emission can be largely explained by an optical depth effect, with the dust continuum being optically thin and the CO lines being optically thick. The effect of radial drift can be seen in the sharpness of the continuum outer edge.

\begin{figure}
\center
\includegraphics[width=.57\textwidth]{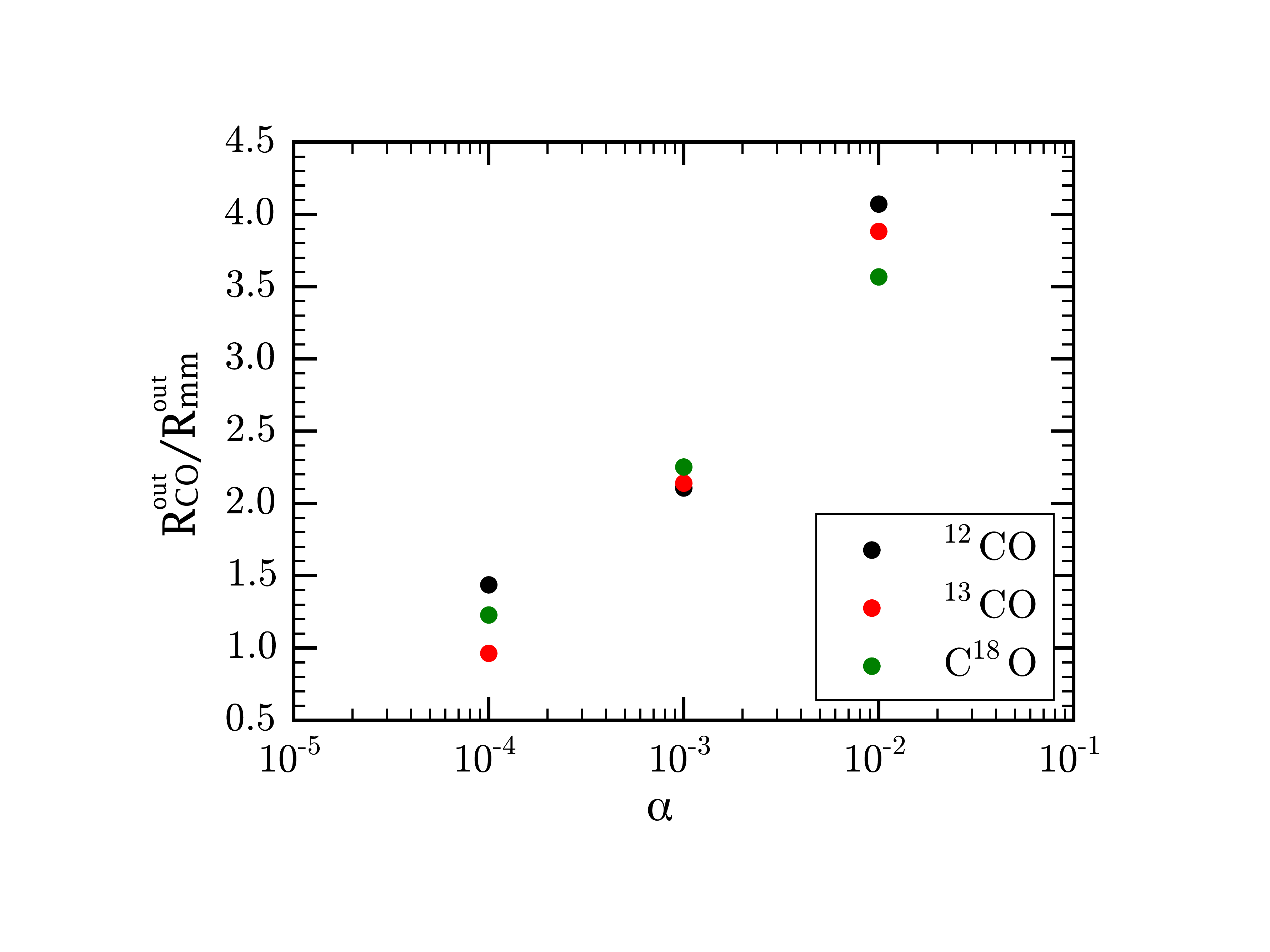}
\caption{Ratio of the CO outer radius $R_{\rm CO}^{\rm out}$ and the mm outer radius $R_{\rm mm}^{\rm out}$ for the different turbulent models, where a dynamic range of 30 has been used for both the continuum and the gas outer radius estimates. Figure adapted from \citet{2017arXiv170506235F}. The color version of the figure is available in the electronic form or in the original paper \citep{2017arXiv170506235F}.
}
\label{fig:rmm_rco}
\end{figure}

Figure \ref{fig:rmm_rco} shows a summary plot of the ratio of disk outer radii as probed by submm continuum ($R_{\rm mm}^{\rm out}$) and CO isotopologue lines ($R_{\rm CO}^{\rm out}$), where we stress again that the underlying gas surface density is exactly the same for the three turbulent models. The ratio increases with turbulence, since $R_{\rm mm}^{\rm out}$ is not significantly affected by the value of $\alpha$, whereas $R_{\rm CO}^{\rm out}$ shrinks for low turbulence values. This result emphasizes how important the simultaneous modeling of both gas and dust can be when interpreting observables as the disk outer radius, and in particular the significance of a proper implementation of realistic grain size distributions in the thermo-chemical models. This holds true for ``simple'' observables as the disk outer radius, but is at least as much important when trying to infer gas surface densities of disks, in particular when we know from observations that the dust properties are changing dramatically with radius, showing a plethora of substructures, as rings, gaps, spirals and azimuthal asymmetries \citep[e.g.][]{2013Sci...340.1199V,2015ApJ...808L...3A,2016Sci...353.1519P}. For example, determining the level of thermal coupling within the dust gaps will be important to correctly interpret the gaps that observations are starting to probe in gas tracers \citep{2016PhRvL.117y1101I,2017ApJ...835..228T,2017A&A...600A..72F}. Good gas temperature estimates across the gaps, e.g. via line ratio determinations, will be important to disentangle between surface density and temperature gaps \citep{facchini17b}.

%\section{The significance of thermal (de)coupling in gaps opened by planets}

%\newpage

\end{document}